\documentclass[preprint,aps]{revtex4}

\newcommand{\beq}{\begin{equation}}
\newcommand\eeq{\end{equation}}
\newcommand{\baa}{\begin{eqnarray}}
\newcommand\ea{\end{eqnarray}}
\newcommand{\pv}[1]{{-  \hspace {-4.0mm} #1}}

\def\nn{\nonumber \\}

\begin{document}
\title{Matrix-model dualities in the collective field formulation}
\author{Ivan Andri\' c and Danijel Jurman
\footnote{e-mail address: 
iandric@irb.hr \\ \hspace*{2.2cm}
djurman@irb.hr }}
\affiliation{Theoretical Physics Division,\\
Rudjer Bo\v skovi\'c Institute, P.O. Box 180,\\
HR-10002 Zagreb, CROATIA}

\begin{abstract}
We establish a strong-weak coupling duality between two types of free matrix models.
In the large-N limit, the real-symmetric matrix model is dual to the quaternionic-real matrix model.
Using the large-N conformal invariant collective field formulation, the duality is displayed in terms 
of the generators of the conformal group.  The conformally invariant master Hamiltonian is constructed 
and we conjecture that the master Hamiltonian corresponds to the hermitian matrix model.\\
\\  
\end{abstract}
\maketitle

\section{Introduction}

Single hermitian $N\times N$ matrix quantum mechanics is connected 
with $1+1$ dimensional strings via 
the double-scaling limit \cite{Group,Das:1990ka,Polchinski:1990mf}.
Recent progress in this field reinterpreted matrix quantum mechanics as a
theory of $N$ D0-branes \cite{McGreevy:2003kb,Klebanov:2003km}.

Two approaches in which matrix models are analysed provide two interpretations. 
In the first approach, the $N$ eigenvalues of a matrix are treated as 
fermions moving classically in the inverted harmonic oscillator
potential \cite{Polchinski:1991uq}.
A single eigenvalue excitation over the filled Fermi sea has been interpreted as a D0-brane
and the matrix degrees of freedom have been interpreted as an open  
tachyon field on the D0-branes.
In the second approach,  the bosonization of fermions is performed by introducing
collective field which 
represents the density of eigenvalues \cite{Das:1990ka,Jevicki:1979mb}.
In this picture, the collective field describes the closed string degrees of freedom
and small deformations of the Fermi sea are described by excitations of 
the massless scalar particle.
An important phenomenon occurs: the space of eigenvalues provides a new 
space dimension in which the string moves, and a holographic description arises \cite{'tHooft:1993gx,Susskind:1994vu}.
Since closed strings are part of gravity theory, the duality between the $0+1$ 
non-gravitational theory (open strings) and the $1+1$ theory which contains
gravity (closed strings) is holographic.
Owing to the correspondence we
expect that calculations in string theory can be performed in the simpler matrix theory.
Such a programme has already been used to study cosmology \cite{Karczmarek:2003pv} and  
particle production in cosmology \cite{Das:2004hw}.
There are attempts to treat black holes in a similar way \cite{Sen:2004yv}.

Correspondence between various matrix models and string theories goes beyond the
hermitian matrix model. It has been shown that unoriented string theories correspond
to real-symmetric and quaternionic-real matrix models, depending on the type
of the orientifold projection under study \cite{Gomis:2003vi}.
Analysing themodynamical propeties of these two models in the 
aforementioned first approach, the duality has been established. 
The same type of duality has emerged in the study of the Calogero-Moser models 
\cite{Andric:2000cn}
to which matrix models reduce in the first approach.

In this paper we show the 
appearance of duality between the real-symmetric and the
quaternionic-real matrix model in the second approach. The first model is invariant to the
$SO(N)$ group and the second to the $Sp(N)$ group of transformations.
The paper is organised as follows.
In section II we develop a general formalism to express a particular matrix model in the collective
field formalism.
In section III we discuss the invariance of the collective field Lagrangian descending from
the matrix model.
Duality relations are displayed in terms of the generators of the conformal symmetry group 
and the master Hamiltonian for dual systems is constructed preserving conformal invariance. 
In section IV, we consider a connection of the master Hamiltonian with the Hamiltonian of the 
hermitian matrix model.
In conclusion we summarize the main results. 

\section{Matrix model and the collective-field Hamiltonian}
The dynamics of the one-matrix model is defined by the action
Ref. \cite{Brezin:1977sv,Group}
\begin{equation}\label{action}
S=\int dt \left (\frac{1}{4} Tr \dot{M}^2(t) - V(M) \right)\;\;,
\end{equation}
with the matrix $M$ of the form
\begin{equation}\label{decom}
M=\sum_{\alpha=0}^{3}m_{\alpha}\otimes e^{(\alpha)}\;\;,\;\alpha=0,1,2,3\;\;,
\end{equation}
where real, $N\times N$ matrices $m_\alpha$'s have the following properties:
\begin{equation}\label{symdec}
m_{0}^{T}=m_{0}\;\;,\;m_{l}^{T}=-m_{l}\;\;,\;l=1,2,3\;\;,
\end{equation}
and elementary quaternions $e^{(\alpha)}$'s are  represented by $2\times 2$ matrices \cite{Mehta}
\begin{equation}\label{eesdef}
e^{(0)}=\left[ \begin{array}r
 	1\; 0\\
0\; 1
\end{array} \right]
\;\;,\;
e^{(1)}=\left[ \begin{array}r
 	i\;\;\;\; 0\\
\;0\; -i
\end{array} \right]\;\;,\;
e^{(2)}=\left[ \begin{array}l
 	0\; -1\\
1\;\;\;\;\; 0
\end{array} \right]
\;\;,\;
e^{(3)}=\left[ \begin{array}r
 	0\; -i\\
-i\;\;\; 0
\end{array} \right]
\;\;.
\end{equation}
From the definition (\ref{eesdef}) we obtain commutation relations
\begin{equation}
\left[ e^{(i)}, e^{(j)} \right] =2\epsilon_{ijk}e^{(k)}\;\;,
\end{equation}
where $\epsilon_{ijk}$ is the totally antisymmetric with respect to the permutations 
of the indices and $\epsilon_{123}=1$.

We consider three types of matrices: real-symmetric, hermitian and quaternionic-real.
The factor $1/4$ in the action (\ref{action}) has been introduced  
to make possible a unique treatment of all three models.
With the definitions given above, Eq.(\ref{decom}) represents a quaternionic-real matrix.
Taking $m_{i}=0$ reduces (\ref{action}) to the familiar expression
\begin{equation}
S=\frac{1}{2}\int dt Tr \dot{R}^2\;\;,
\end {equation}
where $R=m_0$ is the real-symmetric matrix and if we take
$m_{1}=m_{2}=0$, 
the expression (\ref{action}) reduces to 
\begin{equation}
S=\frac{1}{2} \int dt Tr \dot{G}^2\;\;,
\end{equation}
where $G=m_0+im_3$ is the hermitian matrix.

To analyse a matrix model in the large-N limit, we introduce the 
collective field variables
\begin{equation}\label{collvark}
	\phi_{k}(t)=\frac{1}{2} Tr e^{-ikM(t)}\;\;.
\end{equation}
This otherwise over-complete set of variables 
becomes complete in the large-N limit and we can express the action (\ref{action}) in terms
of $\phi_{k}(t)$'s. The general procedure is developed in \cite{Jevicki:1979mb} and
for the quaternionic-real model it was done by expansion and
resummation in $k$ \cite{Andric:1982jk}. Here we present a similar method based
on the tensor product of algebras, which can be generalized for other applications.

Expressed in terms of collective field in coordinate space
\begin{equation}\label{collvarx}
\phi(x,t)=\frac{1}{2}\int\frac{dk}{2\pi}e^{ikx} \phi_{k}(t)\;\;, 
\end{equation}
the free matrix Hamiltonian  is
\begin{equation}\label{noherham}
H=\frac{1}{2}\int\int  dxdy \Omega[\phi;x,y] \pi(x) \pi(y) -\frac{i}{2}
\int dx \omega[\phi;x] \pi(x)\;\;.
\end{equation}
Whereas $\pi(x)$ in (\ref{noherham}) is the canonical conjugate of $\phi(x)$,
 $\Omega[\phi;x,y]$ and $\omega[\phi;x]$ are to be determined by transformation from quantum 
mechanics to collective field theory. 
The factor $1/2$ in the definitions (\ref{collvark}) and (\ref{collvarx}) is present because
of the normalization condition
\begin{equation}\label{constraint}
\int dx \phi(x)=N\;\;,
\end{equation} 
where $N$ is the number of independent eigenvalues of the matrix $M$. Again, the expressions 
(\ref{collvark}) and (\ref{collvarx}) reduce to familiar expressions without the factor $1/2$
in the cases of symmetric and hermitian matrices.

In order to formulate collective field theory for a matrix model, we have
to calculate $\Omega[\phi;x,y]$ and $\omega[\phi;x]$. For this
purpose, we establish some preliminary identities.   
First we notice that if $M$ is a quaternionic-real (real-symmetric, hermitian) matrix, then $M^n$ is 
also a quaternionic-real (real-symmetric, hermitian) matrix
\begin{equation}\label{potm}
 M^{n}\equiv \sum_{\alpha=0}^{3} m_{\alpha}(n)\otimes e^{(\alpha)}\;\;,\; 
m_{0}^{T}(n)=m_{0}(n)\;\;,\;m_{l}^{T}(n)=-m_{l}(n)\;\;,\;l=1,2,3\;\;.
\end{equation}
This statement is easily proved by induction, collecting appropriate terms in 
\begin{equation}\label{stepind}
M^{n+1}=\frac{1}{2}(M^{n}M+MM^{n})\;\;.
\end{equation}
Defining the decomposition of the matrix $exp(-isM)$ in terms of quaternions
\begin{equation} \label{symdecexp}
[s]\equiv e^{-isM} \equiv \sum_{\alpha=0}^{3}[s]_{\alpha} \otimes e^{(\alpha)}
\end{equation}
and using (\ref{potm}) we conclude that matrices $[s]_{\alpha}$'s have the following
properties:
\begin{equation}
[s]_{0}^{T}=[s]_{0}\;\;,\;[s]_{l}^{T}=-[s]_{l}\;\;,\;l=1,2,3\;\;.
\end{equation}   
Now we introduce further decomposition of the symmetric and antisymmetric matrices
in (\ref{symdec}):
\baa \label{finesymdec}
&&m_{0}=\sum_{i,j=1, i\leq j}^{N}m_{0}^{ij}h_{ij}^{+}\;\;,\nn
&&m_{l}=\sum_{i,j=1,i<j}^{N} m_{l}^{ij}h_{ij}^{-}\;\;,\;l=1,2,3\;\;,
\ea
where $h_{ij}^{\pm}$ are elementary matrices with elements at the m-th row and the n-th column defined by
\begin{equation}\label{defhij}
[h_{ij}^{\pm}]_{mn}=\delta_{im}\delta_{jn}\pm \delta_{in}\delta_{jm}\;\;.
\end{equation}
From the definitions of $h_{ij}^{\pm}$  we obtain the following trace rules:
\baa\label{traceid}
&&\sum_{i,j=1}^{N} Tr(Xh_{ij}^{\pm}X'h_{ij}^{\pm})=
2[Tr(XX'^{T})\pm(TrX)(TrX')]\;\;,\nn
&&\sum_{i,j=1}^{N} Tr(Xh_{ij}^{\pm})Tr(X'h_{ij}^{\pm})=
2 Tr (XX'^{T}\pm XX')\;\;.
\ea

After these preliminary remarks  we are in a position to present
the calculation of the relevant
functionals $\Omega[\phi;x,y]$ and $\omega[\phi;x]$.
Passing from the Lagrangian formulation to the Hamiltonian, after performing 
quantization, the transformation to the collective field 
Hamiltonian is obtained by application of the chain rule
\begin{equation}\label{lancanopravilo}
\frac{\partial}{\partial m_\alpha^{ij}}\rightarrow \int dx \frac{\partial \phi(x)}{\partial m_\alpha^{ij}} 
\frac{\delta}{\delta \phi(x)}\;\;.
\end{equation}
Using (\ref{lancanopravilo}), for $\Omega[\phi;x,y]$ and $\omega[\phi;x]$ in (\ref{noherham}) we find
\begin{equation}\label{omega1callc}
\Omega[\phi;x,y]
=\frac{1}{2}\left[\sum_{i,j=1,i\leq j}^{N}
(1+\delta_{ij})
\frac{\partial\phi(x)}{\partial m^{ij}_{0}}
\frac{\partial\phi(y)}{\partial m^{ij}_{0}}+
\sum_{l=1}^{3}\sum_{i,j=1,i<j}^{N} 
\frac{\partial\phi(x)}{\partial m^{ij}_{l}}
\frac{\partial\phi(y)}{\partial m^{ij}_{l}}\right]\;\;,
\end{equation}
\begin{equation}\label{om2call}
\omega[\phi;x]=-\frac{1}{2}\left(\sum_{i,j=1,i\leq j}^{N}(1+\delta_{ij})
\frac{\partial^{\;^2}\phi(x)}{\partial  m^{ij\;^{2}}_{0} }+
\sum_{l=1}^{3}\sum_{i,j=1,i<j}^{N}
\frac{\partial^{\;^{2}}\phi(x)}{\partial m^{ij\;^{2}}_{l}}
\right)\;\;.
\end{equation}
Performing the first summation in (\ref{omega1callc}) we obtain
\baa \label{om1callpart1}
&&\frac{1}{2}\sum_{i,j=1,i\leq j}^{N}
(1+\delta_{ij})
\frac{\partial\phi(x)}{\partial m^{ij}_{0}}
\frac{\partial\phi(y)}{\partial m^{ij}_{0}}=
\frac{1}{8}\int\int\frac{dkdk'}{(2\pi)^{2}}(\partial_{x} e^{ikx})
(\partial_{y} e^{ik'y})\times\nn
&&\times
\sum_{i,j=1,i\leq j}^{N}(1+\delta_{ij})
Tr\left\{[k](1-\frac{1}{2} \delta_{ij})(h_{ij}^{+}\otimes e_{0}) \right\}
Tr\left\{[k'](1-\frac{1}{2} \delta_{ij})(h_{ij}^{+}\otimes e_{0}) \right\}=\nn
&&=\frac{1}{4}\int\int\frac{dkdk'}{(2\pi)^{2}}(\partial_{x} e^{ikx})
(\partial_{y} e^{ik'y})\sum_{i,j=1}^{N}
Tr\left\{[k]_{0}h_{ij}^{+}\right\}Tr\left\{[k']_{0}h_{ij}^{+}\right\}=\nn
&&=\int\int\frac{dkdk'}{(2\pi)^{2}}(\partial_{x} e^{ikx})
(\partial_{y} e^{ik'y})
Tr\left\{ [k]_{0}[k']_{0}\right\}
\ea
and analogously for the other sums in (\ref{omega1callc}) 
\begin{equation}\label{om1callpart2}
 \frac{1}{2}\sum_{i,j=1,i<j}^{N}
\frac{\partial\phi(x)}{\partial m^{ij}_{l}}
\frac{\partial\phi(y)}{\partial m^{ij}_{l}}=
-\int\int\frac{dkdk'}{(2\pi)^{2}}(\partial_{x} e^{ikx})
(\partial_{y} e^{ik'y})
Tr \left\{[k]_{l}[k']_{l}\right\}\;\;.
\end{equation}
The first step in (\ref{om1callpart1}) is obtained from the definitions (\ref{collvark}) and (\ref{collvarx}),
and we have only rewritten
the multiplications by $k$ and $k'$ as appropriate derivatives. The second step is 
obtained by use of the trace property for the tensor product of matrices
\begin{equation}\label{tracetensor}
Tr[(A \otimes a)(B \otimes b)]=Tr(AB) Tr(ab)
\end{equation}
and by the orthogonality of $e^{(\alpha)}$'s:
\begin{equation}\label{ortees}
Tr e^{(\alpha)} e^{(\beta)}=2 \eta^{\alpha \beta}\;\;,\;
\eta=\left[ \begin{array}l
1\;\;\;\;\; 0\;\;\;\;\; 0\;\;\;\;\; 0\\
0\; -1\;\;\;\;\; 0\;\;\;\;\; 0\\
0\;\;\;\;\; 0\; -1\;\;\;\;\; 0\\
0\;\;\;\;\; 0\;\;\;\;\; 0\; -1
\end{array} \right]\;\;.
\end{equation} 
The third, last step in (\ref{om1callpart1}) is obtained by use of
 (\ref{traceid}), (\ref{potm}) and by rewriting the
sum over $i\leq j$ as the sum over $i\neq j$.
Collecting the partial results (\ref{om1callpart1}) and (\ref{om1callpart2})
 we obtain for $\Omega[\phi;x,y]$ 
\begin{equation}\label{om1final}
\Omega[\phi;x,y]=\frac{1}{2}\partial^2_{xy} \int \frac{dk dk'}{(2\pi)^2} e^{ikx}
e^{ik'y}  Tr e^{-i(k+k')M}\;\;.
\end{equation}
The calculation of $\omega[\phi;x]$ is performed in a similar way once the second 
derivatives have been rewritten in a suitable form, using the identity 
\begin{equation}\label{feyiden}
\partial^2 Tr e^{-ikM}=-k^2 \int_0^1 d \beta 
Tr \left[ e^{-ik\beta M} (\partial M) e^{-ik(1-\beta)M} (\partial M) \right]\;\;.
\end{equation} 
As an example, for the first sum in (\ref{om2call}) we have
\baa \label{om2calpart1}
&&-\frac{1}{2}\sum_{i,j=1,i\leq j}^{N}(1+\delta_{ij})
\frac{\partial^{\;^{2}}\phi(x)}{\partial m^{ij\;^{2}}_{0}}=
\frac{1}{4}\int\int^{1}_{0} \frac{dkd\beta}{2\pi} k^{2} e^{ikx} \times \nn
&&\times \sum_{i,j=1,i \leq j}^{N}(1+\delta_{ij})
Tr\left\{ e^{-ik\beta M}\frac{\partial M}{\partial m^{ij}_{0}}
e^{-ik(1-\beta)M}\frac{\partial M}{\partial m^{ij}_{0}} \right\}\;\;.
\ea
From this point the calculation is lenghtly but straigthforward and
the final result is
\begin{equation}\label{om2final}
\omega[\phi;x]=\frac{1}{2} \int \int^{1}_{0} \frac{dk d\beta}{2\pi} k^2 e^{ikx} \left(
-Tr e^{-ikM}+ Tr e^{-i\beta k M} Tr e^{-i(1-\beta)kM} \right)\;\;. 
\end{equation}
Substituing the inverse of (\ref{collvarx})
\begin{equation}\label{inverscollvarx}
Tr e^{-ikM}=2 \int dx e^{-ikx} \phi(x)
\end{equation}
into (\ref{om1final}) and (\ref{om2final}), we find
\baa \label{om12final}
&&\Omega[\phi;x,y]= \partial_{xy}^{2}\left[ \delta (x-y) \phi (y)\right] \nn
&&\omega[\phi;x]=(\lambda-1) \partial_x^2 \phi(x)+2\lambda \partial_x \phi(x) \pv\int dy \frac{\phi(y)}{x-y}\;\;.
\ea
The parameter $\lambda$ in (\ref{om12final}) determines the number of independent matrix
elements $n_\lambda$  in the case of real-symmetric, hermitian and quaternionic-real
matrices:
\begin{equation}\label{numpar}
n_\lambda =\lambda N(N-1)+N
\end{equation}
and $\lambda=1/2,\;1,\;2,$ respectively. $\lambda$ is called 
the statistical parameter because it enters in the exponent of the integration
measure over matrices and therefore in the exponent of the prefactor
in the wave function \cite{Mehta}. If we exchange two eigenvalues, the wave function
changes its phase by $e^{i\pi\lambda}$.
For $\lambda=1$, the statistics of the matrix eigenvalues are fermionic, 
for $\lambda=0$ (diagonal matrix) bosonic and 
for $\lambda=1/2$ and $\lambda=2$ we have an exclusion type of statistics \cite{Polychronakos:1999sx}. 

Finally, after hermitization \cite{Jevicki:1979mb} of (\ref{noherham}) we obtain the collective field Hamiltonian 
\[
H=\frac{1}{2} \int dx \phi(x)(\partial_{x}\pi)^{2}+\frac{1}{2}\int dx\phi(x)
\left(\frac{\lambda-1}{2}\frac{\partial_{x}\phi(x)}{\phi}+\lambda 
\pv\int dy \frac{\phi(y)}{x-y} \right )^{2}+\]
\begin{equation}\label{herham}
-\mu \int
dx \phi(x)+\int dx \phi(x) V(x)-\frac{\lambda-1}{4}\int dx\partial_x^2\delta(x-y)|_{y=x}
-\frac{\lambda}{2}\pv\int dx \partial_x\frac{1}{x-y}|_{y=x}
\;\;,
\end{equation}
where the term with the Lagrange multipliler $\mu$ has been added because of the constraint (\ref{constraint}).  
The last two terms in (\ref{herham}), which are singular, do not contribute in the leading order
in $N$ \cite{Andric:1985ck}.

\section{Conformal invariance and duality}
In this section we find generators of symmetries of the action defined by the 
Lagrangian density
corresponding to the Hamiltonian (\ref{herham})
\begin{equation}\label{herlag}
{\cal L}(\phi,\dot{\phi})=
\frac{1}{2} \frac{ (\partial_x^{-1}\dot{\phi})^2}{\phi}
-\frac{1}{2} \phi \left[ \frac{(\lambda-1)}{2}\frac{\partial_x \phi}{\phi}
+\lambda \pv\int dy \frac{\phi(y)}{x-y} \right]^2\;\;,
\end{equation}
where $\partial_x^{-1}$ is short for $\int^{x} dy \dot{\phi}(y)$.

However, let us first establish a fundamental property of the 
collective field Lagrangian descending from the
matrix models. In order to formulate string theory, 
we need to analyse the matrix model in the critical potential.
 In the cubic theory (the hermitian matrix model), 
it has been shown that the kinetic term induces 
the harmonic potential \cite{Avan:1991kq,Jevicki:1991yi}
\begin{equation}\label{harfromkin}
\int dx dt \frac{\left( \partial_x^{-1} \dot{\phi}\right)^2}{\phi}= \int dx' dt'
 \left[ \frac{ \left( \partial_{x'}^{-1} \dot{ \phi' } \right)^2 }{\phi'}+
x'^2 \phi'( x', t') \right]
\end{equation}
 through a coordinate reparametrization and field rescaling 
\begin{equation}\label{hiptrans}
x=\frac{x'}{\sinh t'}\;\;, t=\tanh t'\;\;,\phi(x,t)=\phi(x',t') \cosh t'\;\;.
\end{equation}  
Similarly, it can be shown that the second term in the Lagrangian (\ref{herlag}) 
remains invariant and therefore all three matrix models have background independence. 
This property enables us to concentrate the discussion on the free models. 

To display the duality of the matrix models, we need infinitesimal generators
of the symmetry of the action defined by the collective field Lagrangian (\ref{herlag}).
The symmetry transformations are the global conformal
reparametrisations of time which leave the action invariant, but
are not the symmetries of the Lagrangian. 
Therefore,  a Noether theorem is needed with an additional
term owing to the change of the form of the Lagrangian: 
\begin{equation}\label{deltas}
\delta S= S'-S=\int dx'dt' {\cal L}\left[\phi'(x',t'),\dot{\phi}'(x',t')\right]-
\int dxdt {\cal L}\left[\phi(x,t),\dot{\phi}(x,t)\right]=\int dt \frac{dA}{dt}\;\;,
\end{equation}
where A is a functional of $\phi$ and $\dot{\phi}$ to be determined. On the other hand,
the change of the action owing to the infinitesimal symmetry
transformation $\delta \phi$, obtained by use of the Euler-Lagrange
equation of motion, is
\begin{equation}\label{deltal}
\delta S=\int \int dt dx \left( \frac{\delta{\cal L}}{\delta \phi} \delta \phi+
\frac{\delta {\cal L}}{\delta \dot{\phi}} \delta \dot{\phi} \right)=
\int dt \frac{d}{dt}\left( \int dx \frac{ \delta {\cal L}}{\delta \dot{\phi}} \delta \phi\right)\;\;.
\end{equation}
This change of action should be equal to (\ref{deltas}) and for the conserved quantity
we obtain
\begin{equation}\label{concha}
Q=\int dx \frac{ \delta {\cal L}}{\delta \dot{\phi}}\delta \phi-A
\;\;. 
\end{equation}

We show that the action determined by the Lagrangian density (\ref{herlag})
possesses three kinds of symmetry: time translation, scaling and special
conformal tranformation. 
The infinitesimal forms of these transformations are
\begin{equation}\label{inftrant} 
t'=t-\epsilon t^{n}\;,\;
\end{equation}
for $n=0,1,2$, respectively.
Under these transformations, the space coordinate $x$, the field $\phi(x,t)$
and the space-time volume element $dtdx$ transform according to
\baa\label{inftranxfivol}
&&x'=\left( \frac{\partial t'}{\partial t} \right )^{d_{x}}x\;\;,\nn 
&&\phi'(x',t')=\left (\frac{\partial t'}{\partial t} \right)^{d_{\phi}}
 \phi(x,t)\;\;,\nn
&&dx'dt'= \left( \frac{\partial t'}{\partial t} \right )^{d_{x}+1} dxdt\;\;.
\ea
Dimensions are determined to be $d_x=1/2$ and $d_\phi=-1/2$.
Performing the infinitesimal transformation (\ref{inftrant}), from (\ref{inftranxfivol}) we
obtain
\begin{equation}\label{inftenfi}
\delta \phi(x,t)=\phi'(x,t)-\phi(x,t)=
(-d_\phi nt^{n-1}+d_x nt^{n-1}x\partial_x +t^n \partial_t)\phi(x,t)\;\;.
\end{equation}
Introducing 
\begin{equation}\label{pidotficon}
\partial_x \pi=\partial_x \frac{\delta {\cal L}}{\delta \dot{\phi}}=
-\frac{1}{\phi} \partial_x^{-1} \dot{\phi}\;\;,
\end{equation}
we find for the first part of the conserved quantity (\ref{concha}) 
\begin{equation}\label{deltaa}
\int dx \frac{\delta {\cal L}}{\delta \dot{\phi}} \delta \phi
=\int dx \pi (x,t) \delta \phi =-\frac{n}{2}t^{n-1}\int dx x \phi(x)
\partial_x \pi + t^n \int dx \frac {\left(\partial_x^{-1} \dot{\phi} \right)^2}
{\phi}
\end{equation}
and after some calculation, from
the difference of the Lagrangians (\ref{deltas}) we obtain  
\begin{equation}\label{conchar012}
A=-\frac{n(n-1)}{4}\int dx x^2 \phi +\frac{t^n}{2}\int dx {\cal L}\;\;.
\end{equation}
Substituing (\ref{deltaa}) and (\ref{conchar012}) in (\ref{concha}) we obtain for $n=0,1,2$
\baa
&&Q_0=H\equiv Q_T\;\;,\nn
&&Q_1=-\frac{1}{2}\int dx \phi(x)\partial_x \pi(x)+tH\equiv Q_S\;\;,\nn
&&Q_2=\frac{1}{2}\int dx x^2 \phi(x)-t\int dx x \phi(x) \partial_x \pi(x) + \frac{t^2}{2}H\equiv Q_C\;\;.
\ea
These conserved quantities close the algebra of the conformal group in one dimension
with respect to the classical Poisson brackets
\begin{equation}\label{poissconchar}
 \left\{ Q_T,Q_S \right\}_{PB}= Q_T\;\;,\left\{ Q_C,Q_S \right\}_{PB}= -Q_C\;\;,
 \left\{ Q_T,Q_C \right\}_{PB}=2 Q_T\;\;.
\end{equation}
Performing the quantisation, simplifying by taking $t=0$ and 
performing a similarity transformation, we
obtain the generators used in Ref. \cite{Andric:2002sp} 
\baa \label{simalg2}
&&T_+\left[\phi,\lambda\right]=-J^{-1/2}Q_T J^{1/2}=-\frac{1}{2}\int dx \phi(x) (\partial_x \pi(x))^2-\frac{i}{2}\int dx 
\omega[\phi;x] \pi(x)\;\;,\nn
&&T_-= Q_C\;\;,\;\;T_0=iJ^{-1/2}Q_S J^{1/2}=-\frac{1}{2}\left( i \int dx x \phi(x) \partial_x \pi(x)+E_0\right)\;\;,
\ea
where the Jacobian $J$ is determined by $\omega[\phi;x]$  
\begin{equation}\label{omegajaccon}
\omega[\phi;x]=\partial_x \left( \phi(x) \frac{\delta \ln J}{\delta \phi(x)} \right)\;\;.
\end{equation}
The constant $E_0=n_{\lambda}/2$ in (\ref{simalg2})
, where $n_{\lambda}$ is given by (\ref{numpar}),
is  the ground-state energy of the Hamiltonian (\ref{herham}) with 
the additional harmonic interaction $V(x)=\frac{x^2}{2}$
 known to be equivalent to the operator $T_0$ up to the similarity transformation \cite{Andric:2003gv}.
We can interpret $E_0$ as the ground-state energy of the $n_\lambda$ independent
harmonic oscillators.
 
After establishing the representation of the su(1,1) algebra
\begin{equation} \label{su11alg}
\left[T_+, T_-\right]= -2T_0\;\;,\;\;\;\left[T_0, T_\pm \right]= \pm 2T_\pm 
\end{equation}
we summarise some known results. 
It has been shown in Ref. \cite{Andric:2002sp} that the eigenfunctionals of the
Hamiltonian (\ref{herham}) can be determined if the zero-energy
eigenfunctionals are known
\begin{equation}
T_+[\phi;\lambda] P_m[\phi]=0\;\;,\;\;\;T_0[\phi;\lambda] P_m [\phi]=\mu_m P_m[\phi]\;\;.
\end{equation}
The functional $J^{1/2}P_m$ is then the zero-energy eigenfunctional in accordance with the conformal invariance
of the Lagrangian. 
Owing to the spectrum generating algebra (\ref{su11alg}) the eigenfunctional of the energy $E$
is the coherent state of the Barut-Girardello-type \cite{barGir}.
In order to show this, we define the operator 
\begin{equation}
\hat{T}=-T_{-}  \frac{1}{T_0+\mu_c}\;\;, 
\end{equation}
which has the canonical commutation relation with $T_+$
\begin{equation}
\left[ T_+,\hat{T} \right]=1\;\;,
\end{equation}
and $\mu_c$ is determined by the eigenvalue of the Casimir operator $\hat{C}$
\begin{equation}
\mu_c=-\frac{1}{2}+\sqrt{\frac{1}{4}-\hat{C}}\;\;,
\end{equation}
\begin{equation}
\hat{C}=T_- T_+ +T_0(T_0+1)\;\;.
\end{equation}
Then, the coherent state 
$e^{E\hat{T}} P_m[\phi]$ 
is the eigenfunctional of $T_+$, and $J^{1/2}e^{E\hat{T}} P_m[\phi]$ of the Hamiltonian (\ref{herham}).
By applying $e^{E\hat{T}}$ to $P_m[\phi]$ and using (\ref{su11alg}), we obtain another form for continuum 
states with the eigenvalue $E$
\begin{equation}\label{solution}
e^{E\hat{T}}P_m[\phi] \sim T_-^{-\left(m+E_0-3/2\right)/2} Z_{m+E_0-3/2}(2\sqrt{ET_-})P_m[\phi]\;\;,
\end{equation}
where $Z_{\alpha}(x)$ stands for the Bessel function.

In addition to (\ref{solution})
there exist other solutions \cite{Jevicki:1991yi,Polychronakos:1994xg,Andric:1994nc}
in the case $\lambda\neq 1$ in the non-perturbative sector of the theory. They are soliton
solutions and appear in the BPS limit, because the kinetic term in the Hamiltonian is of order
$1/N$ with respect to the positive definite second term in (\ref{herham}). The BPS limit leads
to the  first-order integro-differential equation and the solution describes the static tachyonic background
with the energy proportional to the charge of the soliton. The non-BPS solutions are obtained from the 
Heisenberg equations of motions. These are moving soliton solutions \cite{Andric:1994nc,Jevicki:1991yi,Polychronakos:1994xg} and for 
$\lambda=1/2$, they describe holes in the background (condensate) and for $\lambda=2$, lumps above the
background. 

The solitons of the collective field description are dual quasi-particles.
To display this duality, we introduce a new field $m(x,t)$ describing
quasi-particles, and this new field will enter in the prefactor of the wave functional.
For the prefactor we take the continuum analogue of the prefactor used in the discrete case \cite{Andric:2000cn}:
\begin{equation}\label{prefactor}
V^{\kappa}[\phi,m]=e^{\kappa \int \int dxdy \phi(x) \ln |x-y| m(y)}\;\;. 
\end{equation}
The duality is displayed by the following relations:
\baa \label{dual1}
&&T_+[\phi,\lambda] V^{\kappa}[\phi,m]=\left[ -\frac{\lambda}{\kappa} T_{+} [m,\kappa^2/\lambda]+
\frac{\kappa \pi^2}{2}\int dx \phi(x) m(x) (\lambda \phi(x)+\kappa m(x))+ \right.\nn
&&\left.+
\frac{(\lambda+\kappa)(\kappa-1)}{4}\int \int dxdz\frac{m(z)\partial_x \phi(x)-\phi(x)\partial_z m(z)}{x-z}\right]
 V^{\kappa}[\phi,m]\;\;,
\ea
\begin{equation}\label{dual2}
T_0[\phi,\lambda]V^{\kappa}[\phi,m]=-\left( T_0[m,\kappa^2/\lambda]+
\frac{E_0(N,\lambda)+E_0(M,\kappa^2/\lambda)+\kappa NM}{2}\right)V^{\kappa}[\phi,m]\;\;.
\end{equation}
Here we have the manifest strong/weak coupling duality. If we interchange the fields $\phi(x)$ and $m(z)$,
the coupling constant
$\lambda$ goes to $\kappa^2/\lambda$. The duality relations (\ref{dual1}) and (\ref{dual2}) are crucial. They
enable us to construct new su(1,1) generators for the system of particles and dual quasi-particles
\[{\cal T}_+=T_+[\phi,\lambda]+\frac{\lambda}{\kappa}T_+[m,\kappa/\lambda]+{\cal H}_{int}\;\;,\]
\[{\cal H}_{int}=
-\frac{(\lambda+\kappa)(\kappa-1)}{4}
\int \int dxdz\frac{m(z)\partial_x \phi(x)-\phi(x)\partial_z m(z)}{x-z}+\]
\[-\frac{\kappa \pi^2}{2}\int dx \phi(x) m(x) (\lambda \phi(x)+\kappa m(x))\;\;,
\]
\[ 
{\cal T}_{0}=T_{0}[\phi,\lambda]+T_{0}[m,\kappa^2/\lambda]\;\;,\]
\begin{equation} \label{masalg}
{\cal T}_{-}=T_{-}[\phi,\lambda]+\frac{\kappa}{\lambda} T_{-}[m,\kappa^2/\lambda]\;\;.
\end{equation}
We interpret the operator ${\cal T}_{+}$ in (\ref{masalg}) as a non-hermitian Hamiltonian for the system
of particles and quasi-particles.
After hermitisation of ${\cal T}_{+}$ we obtain the hermitian form  
\begin{equation}\label{masham}
{\cal H}^M=H[\phi,\lambda]+\frac{\lambda}{\kappa}H[m,\kappa^2/\lambda]+{\cal H}_{int},
\end{equation} 
where $H[\phi,\lambda]$ is the Hamiltonian (\ref{herham}) and $H[m,\kappa^2/\lambda]$ is obtained from 
(\ref{herham}) by substituting $m$ for $\phi$ and $\kappa^2/\lambda$ for $\lambda$.
\section{Master Hamiltonian and the hermitian matrix model}
In this section we argue that the master Hamiltonian (\ref{masham}) corresponds to the hermitian 
matrix model.
The starting point is the Lagrangian   
\begin{equation}\label{hermatlag}
L=\frac{1}{2} Tr \dot{G}^2\;\;,
\end{equation}
where $G$ is a hermitian matrix.
Following the usual procedure of quantization, from the Lagrangian (\ref{hermatlag}) we obtain the Hamiltonian
in coordinate representation
\begin{equation}\label{haha}
H_G=-\frac{1}{2}\partial_G^2\equiv
-\frac{1}{2} \sum_{i,j=1,i\leq j}^{N}
(1+\delta_{ij})
\frac{\partial}{\partial g^{ij}}
\frac{\partial}{\partial g^{ij}}\;\;,
\end{equation}
where $g^{ij}$'s are elements of the matrix $G$.   
Expressed in terms of the collective field,  the Hamiltonian (\ref{haha}) corresponds 
to the collective Hamiltonian (\ref{herham}) with
$\lambda=1$.

Next, we decompose the hermitian matrix into the sum
of the symmetric and antisymmetric matrix, $G=R+iA$.
This decomposition is followed by the decomposition of the  
Hamiltonian (\ref{haha})   
\begin{equation}
H_G= -\frac{1}{2}\partial_G^2=-\frac{1}{2}\partial_R^2+\frac{1}{2}\partial_A^2\;\;.
\end{equation}
Now, if we restrict the wavefunction of the system to be dependent only 
on the eigenvalues of the symmetric matrix, by choosing collective field variables
\begin{equation}\label{colvarsim}
\phi_{k}=Tr e^{-ikR}\;\;,
\end{equation}
we obtain the collective Hamiltonian (\ref{herham}) with the parameter $\lambda=1/2$. 
Using duality relation (\ref{dual1}) for $\kappa=1$, 
we construct the master Hamiltonian (\ref{masham}) 
which contains the following:\\
the Hamiltonian (\ref{herham}) with parameter $\lambda=1/2$ 
describing symmetric-matrix degrees of freedom, 
the Hamiltonian (\ref{herham}) with parameter $\lambda=2$ 
describing quaternionic-matrix degrees of freedom
and the Hamiltonian of the interaction between the field $\phi(x)$ and dual field $m(x)$.

As the next step, we argue that the Hamiltonian (\ref{herham}) with $\lambda=2$ describes a system with
antisymmetric-matrix degrees of freedom. 
To show this, we have to find appropriate collective field variables
for the antisymmetric matrix model.
By naive generalisation, we might use 
\begin{equation}\label{wrong}
m_k=Tr e^{-ikA}\;\;,
\end{equation}
but this choice is misleading. As a first objection, we recall that
the antisymmetric matrix has pure imaginary eigenvalues implying exponential
behaviour of the variables (\ref{wrong}).
As a second objection, we notice that
$A^n$ is not antisymmetric matrix and this property implies that the set of elementary matrices
from the decomposition of $A$ is not complete.
To resolve this puzzle, we use isomorphism between antisymmetric and
quaternionic-real matrices.
We write the antisymmetric matrix
 ($2n \times 2n$) in terms of quaternions:
\begin{equation}
 A=A_{2}\otimes e^{(0)}-iA_{3}\otimes e^{(1)}-R \otimes e^{(2)}+iA_{1} \otimes e^{(3)}\;\;,
\end{equation}
where
\begin{equation}
R^T=R\;\;,\;\;\;A^T_{l}=-A_{l}
\end{equation}
and define the quaternionic-real matrix
\begin{equation}\label{antquat}
Q=\frac{1}{2}\left(\{e^{(2)},A\}+i[e^{(2)},A]\right)=
R \otimes e^{(0)}+A_{1}\otimes e^{(1)}+A_{2}\otimes e^{(2)}+A_{3}\otimes e^{(3)}\;\;,
\end{equation}
which has real eigenvalues and for which we know that
$Q^n$ is also the quaternionic-real matrix.
Noticing that 
\begin{equation}
-Tr \dot{A}^2=Tr \dot{Q}^2 \rightarrow -\partial^2_A=\partial^2_Q\;\;,
\end{equation}
then choosing the collective field variables
\begin{equation}\label{right}
m_k=\frac{1}{2} Tr e^{-ikQ}
\end{equation}
and expressing $\partial_Q^2$ in terms of these
we obviously end up with the Hamiltonian (\ref{herham}) for $\lambda=2$.

It follows from the above discussion that starting from 
the Hamiltonian (\ref{hermatlag}) and restricting the
wave function of the system to be dependent only on the
eigenvalues of the symmetric matrix 
and the quaternionic matrix defined by (\ref{antquat}), the corresponding collective field Hamiltonian
is 
\begin{equation}\label{hamhemhm} 
H=H_{1/2}+\frac{1}{2} H_{2}\;\;,
\end{equation}
where $H_{1/2}$ and $H_{2}$ are the collective field Hamiltonians of the symmetric and quaternionic matrices,
respectively.
Notice that the factor $1/2$ in the Lagrangian (\ref{hermatlag}) instead of $1/4$ as in
(\ref{action}) introduces the factor $1/2$ in front of $H_{2}$. Comparing (\ref{hamhemhm}) with (\ref{masham}),
we see that (\ref{hamhemhm}) is equal to the master Hamiltonian without interaction terms.

To show that the master Hamiltonian indeed describes the hermitian matrix model
we have to generate interaction term. This is achieved by extracting  the prefactor
$\Pi_{i,\alpha}(x_i-z_\alpha)$ ($x_i$ and $z_\alpha$ are the eigenvalues of the symmetric and
quaternionic matrices, respectively) from the wavefunction and defining new Hamiltonian
\begin{equation}
\tilde{H}_G=\Pi_{i,\alpha}(x_i-z_\alpha)^{-1}H_G\Pi_{i,\alpha}(x_i-z_\alpha)\;\;.
\end{equation}
This prefactor has been found in \cite{Andric:2000cn} to appear in 
the ground-state and  
it would be interesting to see whether it is connected with the  
integration measure. Physically, this prefactor prevents finding the particle and quasi-particle 
at the same point.
Expressing this new Hamiltonian in terms of the collective-fields (\ref{colvarsim}) and (\ref{right}),
we obtain the master Hamiltonian (\ref{masham}).

Finally, we see that duality relations enabled us 
to construct the master Hamiltonian and 
to recover the degrees of freedom of the hermitian matrix,
which were lost by choosing (\ref{colvarsim}) as the collective field variables.
\section{Conclusion}
We have seen that su(1,1) algebra generates the dynamical symmetry of the matrix models in the collective
field approach. 
This algebra makes possible the construction of eigenfunctionals,
an explicit display of duality relations between matrix models 
and the construction of the master Hamiltonian in a conformally invariant way.      
We conjecture that the master Hamiltonian with $\lambda=1/2$ describes 
the hermitian matrix model.
 This gives deeper insight into 
the properties of the hermitian matrix model.
In ref. \cite{Jevicki:1991yi}, the hermitian matrix model was analysed by the exact
construction of eigenstates represented by the 
Young-tableaux. 
For the Yang-tableaux only with one column and only with one row, 
effective Lagrangians were constructed, which correspond to the $\lambda=1/2$
and $\lambda=2$ Hamiltonians (\ref{herham}) in our language. 
Further analysis of the dynamics of the master Hamiltonian
could give us more information on closed string states described in the collective
field approach.

\section*{ Acknowledgment}
We would like to thank A. Jevicki for discussion and
hospitality at Brown University, where part of this article
has been done. 
We are also thankfull to L. Jonke for discussion
and A. Kapustin for usefull insight. 
This work was supported by the Ministry of Science and Technology of the
Republic of Croatia under contract No. 0098003.

\end{document}